\documentclass[prd,aps,twocolumn,superscriptaddress,nofootinbib,longbibliography]{revtex4-2}

\pdfoutput=1

\usepackage[colorlinks=true
,urlcolor=blue
,anchorcolor=blue
,citecolor=blue
,filecolor=blue
,linkcolor=blue
,menucolor=blue
,linktocpage=true
,pdfproducer=medialab
,pdfa=true
]{hyperref}
\usepackage[T1]{fontenc}
\usepackage{fontawesome}
\usepackage{feynmf}
\usepackage{graphicx}
\usepackage{enumitem}
\usepackage{latexsym}
\usepackage{amsfonts}
\usepackage{amssymb}
\usepackage{mathrsfs}
\usepackage[svgnames,x11names,table]{xcolor}
\usepackage{amsmath}
\usepackage[capitalise]{cleveref}
\usepackage{slashed}
\usepackage{dcolumn}
\usepackage{verbatim}
\usepackage{comment}
\usepackage{float}
\usepackage{multirow}
\usepackage{xspace}
\usepackage[normalem]{ulem}
\usepackage{lipsum}

\newcommand{\GeV}{~\text{GeV}}

\def\triumf{TRIUMF, 4004 Wesbrook Mall, Vancouver, BC V6T 2A3, Canada}

\begin{document}

\title{Low-Multiplicity Jets as Probes of GeV-Scale Light-Quark-Coupled Particles}
\affiliation{\triumf}

\author{Carlos Henrique de Lima}
\email{cdelima@triumf.ca}
\affiliation{\triumf}

\author{David McKeen}
\email{mckeen@triumf.ca}
\affiliation{\triumf}

\author{Maximilian Swiatlowski}
\email{mswiatlowski@triumf.ca}
\affiliation{\triumf}

\begin{abstract}
We propose a search at the LHC for GeV-scale particles coupling predominantly to light quarks based on low-multiplicity jets. The search targets production in association with a hard photon and uses the feature that a light gauge-singlet can only decay into a small number of hadronic channels, yielding jets with anomalously low charged-track multiplicity and mass compared to QCD jets at the same transverse momentum. We determine the sensitivity to scalar and pseudoscalar couplings to up-quarks, and suggest a data-driven estimate that reduces the sensitivity to jet modeling uncertainties. This search extends the reach to hadronically-coupled particles into a previously inaccessible regime.
\end{abstract}

\maketitle

\section{Introduction} \label{sec:int}

The Large Hadron Collider (LHC) has been incredibly successful in expanding our understanding of the Standard Model (SM) and narrowing down what lies beyond it. The exploration of physics beyond the SM at the LHC has largely focused on pushing into the TeV scale, where the combination of high center-of-mass energy and large integrated luminosity provides unparalleled sensitivity to heavy new states. In contrast, the GeV-scale regime remains comparatively less explored at high-energy colliders.

GeV-scale states that couple to first-generation quarks are present in well-motivated SM extensions, e.g., dark matter mediators or explanations of experimental anomalies~\cite{Feng:2016jff,Feng:2016ysn,Kozaczuk:2016nma,Batell:2017kty,Batell:2018fqo,Egana-Ugrinovic:2018znw,Egana-Ugrinovic:2019dqu,Batell:2021xsi,Murayama:2026ioh} and are generically not difficult to produce at the LHC. The challenge lies in identifying their signature. If these states decay promptly to hadrons, their final states are typically assumed to be indistinguishable from the overwhelming Quantum Chromodynamic (QCD) background. Consequently, a broad class of light, quark-coupled particles has received limited scrutiny at the LHC~\cite{Essig:2013lka,Blackstone:2024ouf,CidVidal:2019urm,Alimena:2025kjv}.

In this Letter, we show that a GeV-scale gauge-singlet particle produced at the LHC and decaying hadronically does not resemble the QCD background. While jets at high $p_T$ (transverse momentum) are characterized by substantial radiation and charged-particle multiplicities, the decay of a light resonance is kinematically limited. For invariant masses below a few GeV, only a few exclusive hadronic channels are open, and the decay products populate a restricted region of phase space. As a result, the corresponding jets exhibit anomalously low charged-track multiplicity and mass.

The use of charged-track multiplicity~\cite{ATLAS:2016vxz} as a discriminator is well established in LHC analyses. It is widely used to distinguish quark and gluon jets~\cite{Gallicchio:2011xq,Frye:2017yrw}, in hadronic $\tau$ identification~\cite{ATLAS:2014rzk}, and in searches for heavy boosted resonances~\cite{Thaler:2010tr,CMS:2022tqn}. 

In an analogous approach, we propose a search for GeV-scale particles that couple predominantly to light quarks using low-multiplicity jets in the $j + \gamma$ final state. As a concrete benchmark, we consider a spin-0 particle with scalar or pseudoscalar couplings to the up quark, over the mass range $0.5~\mathrm{GeV} \lesssim m_\phi \lesssim 20~\mathrm{GeV}$. We find that both charged-track multiplicity and charged-track jet mass\footnote{Using only the charged components of the jet creates an observable robust against experimental considerations such as pile-up.} are powerful probes of the existence of these particles.

Although we focus on a scalar coupled to up quarks, the features studied here are generic to any GeV-scale color-singlet decaying to hadrons and produced in a hard process, such as  Higgs-mixed scalars, scalars coupled to other light flavors, gluon-coupled axion-like particles or hadrophilic vectors; only the production rate and the details of the decay change.

\section{GeV-Scale Light-Quark–Coupled New Physics}\label{sec:model}

\begin{figure*}[bth!]
    \centering
\includegraphics[width=0.495\linewidth]{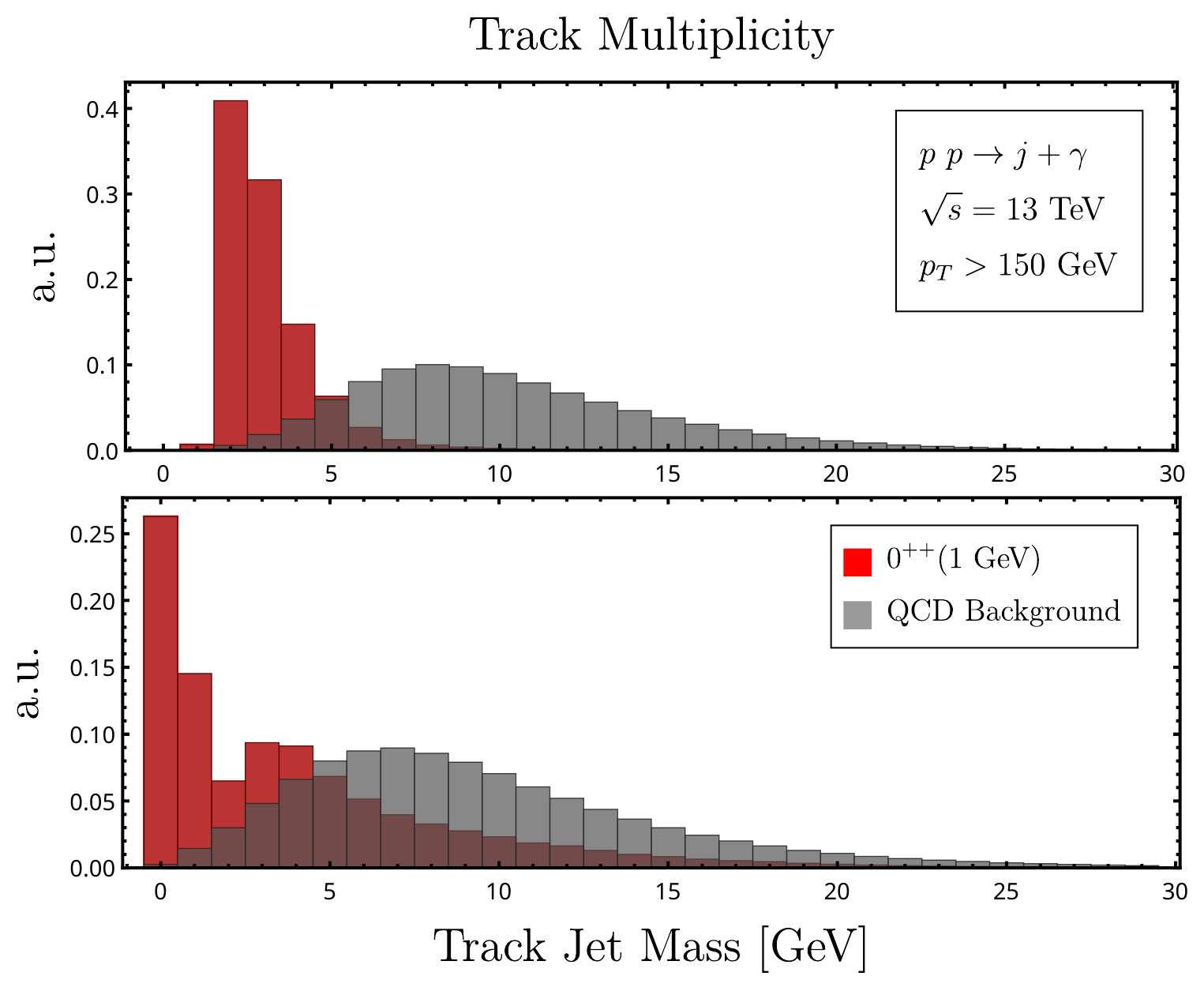}
\includegraphics[width=0.495\linewidth]{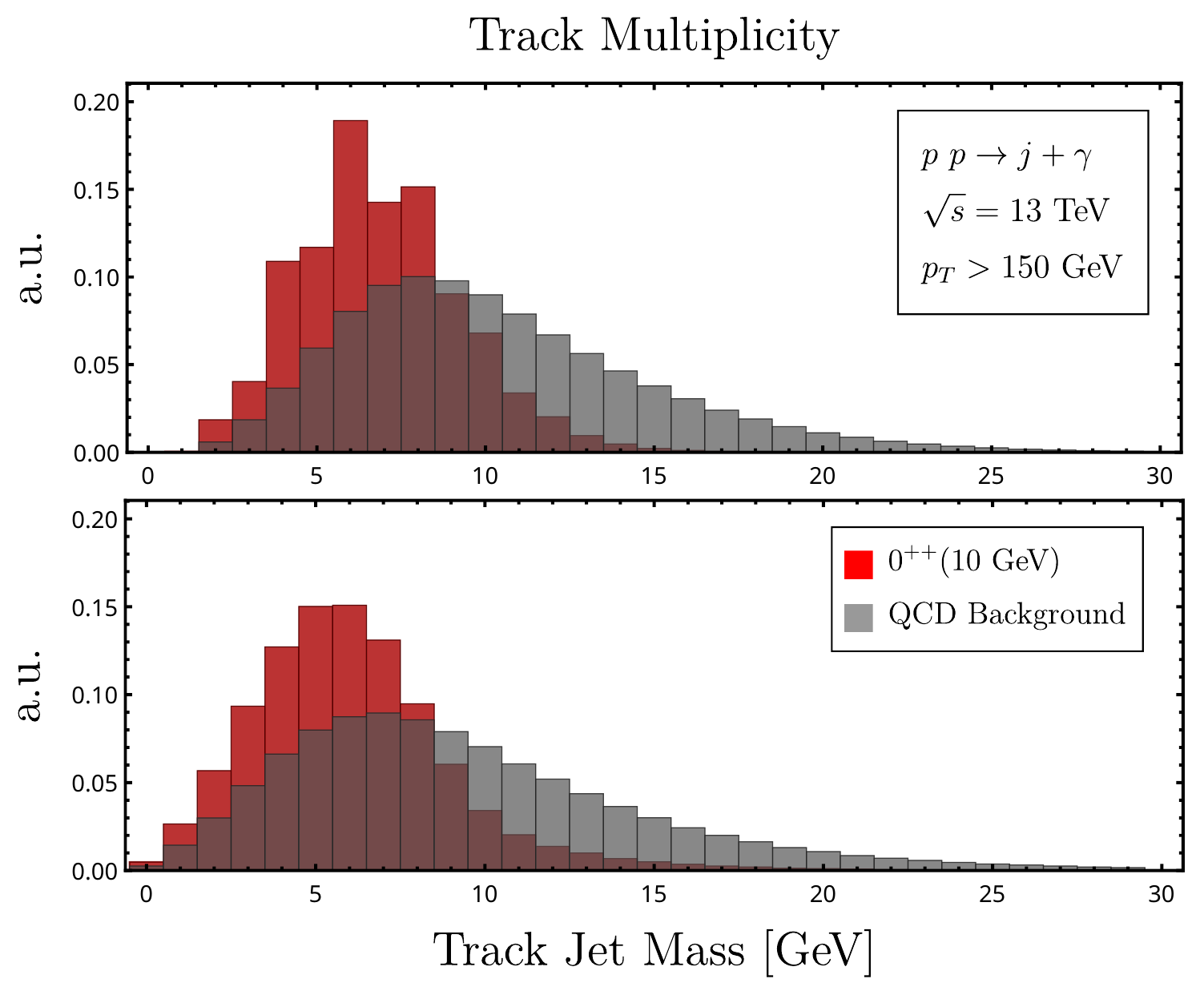}
\caption{Normalized distributions for detector-level Track Multiplicity (\textbf{top}) and Track Jet Mass (\textbf{bottom}) for two mass hypotheses of a $0^{++}$ scalar: 1 GeV (\textbf{left}) and 10 GeV(\textbf{right}). The QCD background from \texttt{MadGraph5\_aMC@NLO} $+$ \texttt{Pythia8.2} is shown in {\color{Gray!10!Black}{gray}}, and the signal in {\color{DarkRed}{red}}. For these plots, we apply the basic kinematic cuts described in the paper.}
\label{fig:mass}
\end{figure*}

Light gauge‑singlet scalars with enhanced couplings to specific SM light flavors provide a compelling portal to new physics at the GeV scale~\cite{Feng:2016jff,Feng:2016ysn,Kozaczuk:2016nma,Batell:2017kty,Batell:2018fqo,Egana-Ugrinovic:2018znw,Egana-Ugrinovic:2019dqu,Batell:2021xsi,Murayama:2026ioh}. Unlike Higgs‑portal scenarios~\cite{McKeen:2008gd,Schmidt-Hoberg:2013hba,Clarke:2013aya}, in which scalar couplings are proportional to fermion masses, a flavor‑specific scalar $\phi$ can couple predominantly to the first generation, e.g., the up quark.\footnote{Similar signatures can be obtained from coupling to other light flavors.} Such a coupling may not induce large flavor-changing neutral currents~\cite{Batell:2017kty,Egana-Ugrinovic:2018znw,Batell:2018fqo,Egana-Ugrinovic:2019dqu,Batell:2021xsi} if some flavor alignment can be arranged (see, e.g., Ref.~\cite{Knapen:2015hia} for a possible mechanism).  At low energies, their interaction is completely determined by their $J^{PC}$ assignment. For a spin-0, we consider the following leading low-energy interactions
\begin{align}\label{eq:eff_coupling}
    \mathcal{L}_{\rm int} &= -\bar{u}\left(\kappa_S S +i\gamma^5\kappa_A A\right)u  \, ,
\end{align}
where $S$ and $A$ are real scalars with $J^{PC}=0^{++}$ and $0^{-+}$ respectively. In our analysis, we consider either $S$ or $A$ existing in isolation, which we generically term $\phi$ when convenient. It is, however, possible that both particles are present together~\cite{Peccei:1986pn,Krauss:1986wx,Murayama:2026ioh}. In the GeV-scale QCD axion model of Ref.~\cite{Murayama:2026ioh}, for example, $S$ and $A$ are the components of a complex scalar with a Peccei--Quinn charge whose vacuum expectation value is induced by the QCD quark condensate; $A$ is identified with the axion, and both states have GeV-scale masses.

We consider the perturbative regime of the scalar self-interaction, where the splitting $\phi\to\phi\phi$ is suppressed relative to the QCD decay and does not alter the jet. Considering also that the couplings that are going to be probed are small, we consider the running induced by the scalar subleading compared to QCD. It is important to note that the couplings are renormalized by QCD, and as we will be probing at LHC energy scales, the running down to GeV can be significant. The couplings in Eq.~(\ref{eq:eff_coupling}) have the same renormalization as the up-quark mass, as $\kappa_{S/A}$ and $m_u$ multiply operators with the same chiral structure~\cite{Larin:1993tq}, so that at one loop
\begin{equation}
\kappa(\mu)=\kappa(\mu_0)\left[\frac{\alpha_s(\mu)}{\alpha_s(\mu_0)}\right]^{12/(33-2N_f)},
\end{equation}
and the ratio $\kappa/m_u$ is renormalization-group invariant. The LHC production measures $\kappa$ at the hard scale of the process, which for our selection is $\mu_{\rm LHC}\simeq150$~GeV, and we quote $\kappa_{S/A}$ at this scale in the $\overline{\rm MS}$ scheme throughout.

The low-energy operators in Eq.~(\ref{eq:eff_coupling}) are not electroweak gauge invariant. Generating these requires the existence of new heavy states with electroweak charges. This operator can arise from $(c/M)\,\phi\,\overline{Q_1}\tilde Hu_R$ after integrating out vector-like quarks or a second Higgs doublet coupled to the first generation at the scale $M$, with $\kappa=cv/(\sqrt2M)$~\cite{Batell:2017kty,Batell:2021xsi}. The $\phi$ mass is then natural for $\kappa(M)\lesssim16\pi^2m_\phi v/(\sqrt2M^2)\simeq0.03\,(m_\phi/{\rm GeV})({\rm TeV}/M)^2$~\cite{Batell:2018fqo}, which with $m_u(1~{\rm TeV})\simeq1.1$~MeV reads $\kappa/m_u\lesssim30\,(m_\phi/{\rm GeV})({\rm TeV}/M)^2~{\rm GeV}^{-1}$, i.e.\ $\kappa(\mu_{\rm LHC})\lesssim0.03\,(m_\phi/{\rm GeV})({\rm TeV}/M)^2$, so that the couplings probed here require mediators around the TeV scale, which can be searched for directly with model-dependent bounds~\cite{Batell:2021xsi}, or a tuning of $m_\phi$. If $A$ is a pseudo-Goldstone boson, $\kappa_A/m_u=1/f$ and the couplings probed here correspond to $f$ of the order of tens of MeV, as in GeV-scale QCD axion models~\cite{Murayama:2026ioh,DiLuzio:2026poh}. We focus on {\em direct} probes of the operators in Eq.~(\ref{eq:eff_coupling}), which are independent of these UV details, to make model-independent statements about the existence of GeV-scale bosons coupled to light quarks.

\section{Low Multiplicity search at the LHC}\label{sec:search}

After fixing the quarks the new boson couples to, the signal depends on three physical properties: its mass, its parity, and charge. After production at the LHC, the (neutral) boson decays to an even number of charged states and additional neutral particles, depending on their parity. Our analysis focuses on the $pp \rightarrow j \gamma$ final state, where the jet either comes from the new scalar particle or from QCD. We consider a scalar $\phi$ which is either P-even or P-odd; however, our search strategy is largely insensitive to this distinction, since both pile-up and underlying events include additional particles, making their jets similar. The signal is $\phi + \gamma$ with $\phi$ emitted from a $u$ quark line and promptly decaying to a few charged tracks. For this channel, we focus on the mass region $0.5~\text{GeV} < m_\phi < 20~\text{GeV}$. Another relevant production channel is $W^\pm + \phi$. This process is more model-dependent due to the sensitivity of the non-gauge-invariant low-energy interaction~\cite{Dror:2017ehi,Dror:2017nsg}. While we do not explore this additional channel in this work, it may provide competitive reach and could be combined with $\gamma + j$ to increase sensitivity.

Since the production is perturbative at the LHC, the only challenging part of signal simulation is the decay of the boson once produced. For a particle with a mass in the GeV range, this can be a formidable task. Current showering and hadronization tools optimized for high energy do not preserve the parity of the $u\bar{u}$ pair, which is largely irrelevant at higher invariant masses. However, for $0.5~\text{GeV} < m_\phi < 1.5~\text{GeV}$, the decay phase space is constrained with only a few final states kinematically available, making the difference between the decays of even- and odd-parity bosons more significant. In our analysis, we use chiral perturbation theory ($\chi$PT) augmented with the interactions of Eq.~(\ref{eq:eff_coupling}) to determine the decay modes up to masses of $1.5~\rm GeV$. We describe our implementation of $\chi$PT in Appendix~\ref{app:ChpTPythia}. For masses above $1.5~\rm GeV$, we use the showering and hadronization of \texttt{Pythia8.2}~\cite{Sjostrand:2014zea} and \texttt{Herwig7.2}~\cite{Bellm:2019zci}.

The distinguishing feature of this signal is the presence of events with anomalously low track multiplicities and jet masses, which are unlikely to arise from QCD alone. It is intrinsically a different kind of jet, induced by a gauge-singlet, which explains why it lies on the low side of track multiplicity and jet mass distributions compared to quark- and gluon-induced jets with the same $p_T$ as seen in Fig.~\ref{fig:mass}.

\begin{figure}[t!]
\centering
\includegraphics[width=1.0\linewidth]{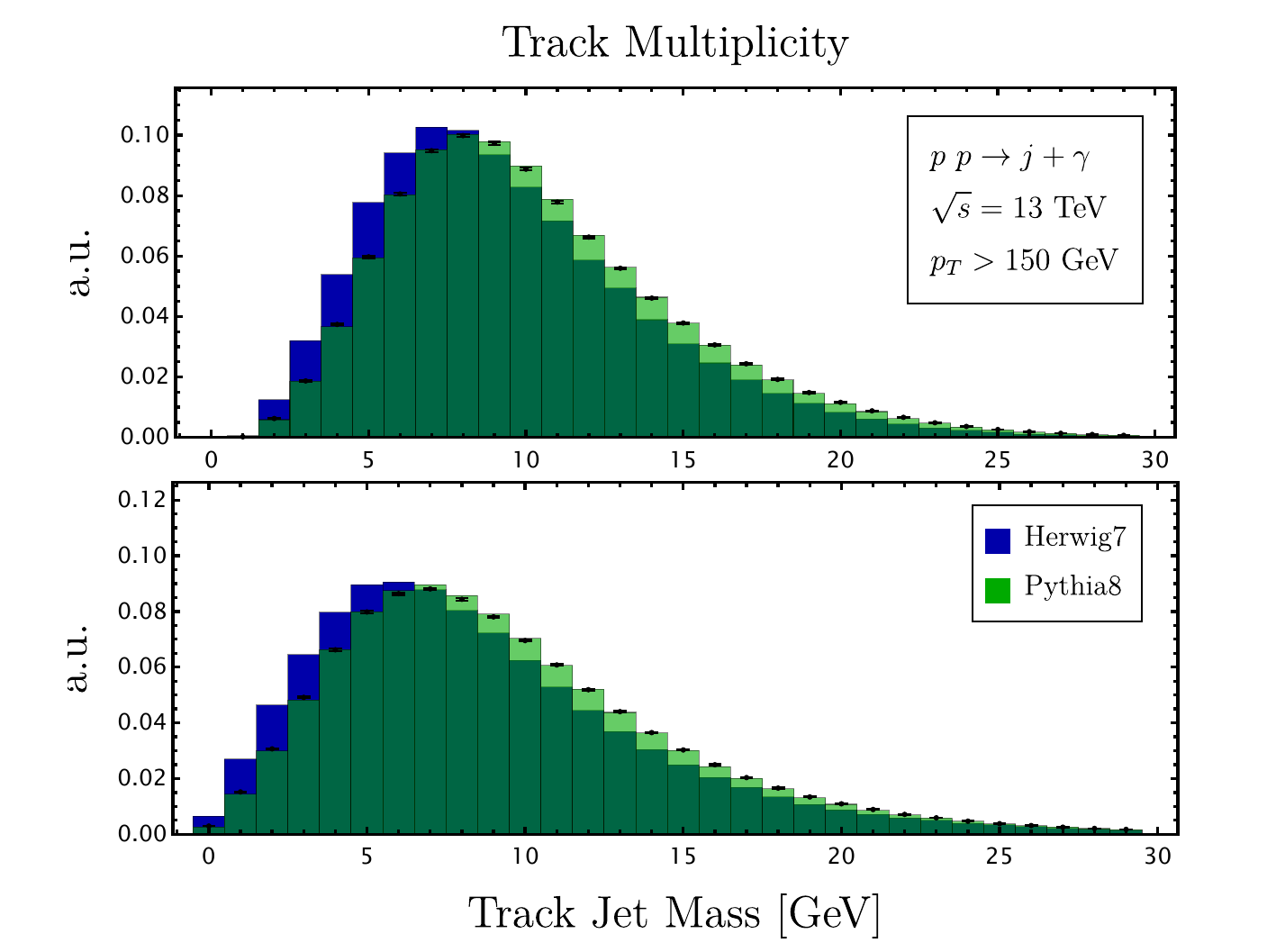}
\caption{ Comparison between the showering and hadronization of a single jet performed in \texttt{Herwig7.2} ({\color{DarkBlue}{blue}}) and \texttt{Pythia8.2} ({\color{DarkGreen}{green}}). Normalized Track Multiplicity (\textbf{top}) and Track Jet Mass (\textbf{bottom}) at the detector level. The black points with error bars show the simulation-to-simulation template estimate of the \texttt{Pythia8.2} jet obtained from the di-jet sample, as discussed in Appendix~\ref{app:datadriven}.}
\label{fig:HWPY}
\end{figure}

The dominant background comes from QCD $j+\gamma$.\footnote{If the analysis were extended to higher scalar masses, it would be necessary also to consider $t\bar{t}$ and electroweak processes as backgrounds.} We simulate the signal and QCD $j + \gamma$ processes at leading order using \texttt{MadGraph5\_aMC@NLO}~\cite{Alwall:2014hca} interfaced with \texttt{Pythia8.2}~\cite{Bierlich:2022pfr,Sjostrand:2014zea} for showering and hadronization. We also cross-check the showering and hadronization using \texttt{Herwig7.2}~\cite{Bahr:2008pv,Bellm:2019zci}, shown in Fig.~\ref{fig:HWPY}. Additionally, our analysis ignores backgrounds from ``fake'' photons where a jet is mis-reconstructed as a photon; these backgrounds are assumed to be small but would need to be estimated in a complete analysis. Jet clustering is done with \texttt{FastJet}~\cite{Cacciari:2011ma} using the anti-$k_T$ algorithm, with $R=0.4$. We use a simplified detector simulation implemented in \texttt{DELPHES}~\cite{deFavereau:2013fsa} using the ATLAS detector card. Pile-up from simultaneous $pp$ collisions is not simulated, but as the main observables use inner detector tracks, which are largely robust to contamination, the impacts of pile-up on the analysis are expected to be small.

While consistency between \texttt{Pythia8.2} and \texttt{Herwig7.2} from Fig.~\ref{fig:HWPY} is encouraging, simulations of extreme jet properties are not necessarily expected to agree with data. To address this, in Appendix~\ref{app:datadriven} we describe a data-driven template method inspired by existing work by LHC collaborations~\cite{Gallicchio:2011xq,Larkoski:2019nwj,ATLAS:2014vax,ATLAS:2023dyu}, and show that it can be used to derive a background estimate with only theoretically well-understood quark and gluon fractions extracted from simulation. This approach, established in LHC quark/gluon tagging analyses, is insensitive to the non-perturbative uncertainties that dominate the showering of low-multiplicity jets, and it would account correctly for the rare low-multiplicity tail from QCD resonances. For our analysis, we use the templates from \texttt{Pythia8.2}, and as the rate is calculated at LO, we include an additional $\pm50\%$ background uncertainty in the statistical inference.

We require that the jet and photon are reconstructed in the main barrel ($\eta<1.4$) or in the endcaps ($1.5<\eta<2.4$). We enforce that both have $p_T > 150~\text{GeV}$ to pass trigger requirements in Run 2 and 3 of the LHC. We veto any process that produces additional jets, leptons, or photons. To slightly improve the reach, we also perform a loose b-tag veto to remove heavy-flavor events from the background. We also consider a lower cut on the jet mass and track jet mass reconstruction based on the detector resolution of 500 MeV. This slightly reduces the reach for light-mass scalars, but pile-up and underlying-event broadening already make it challenging to distinguish different masses and parity hypotheses.

Under these basic cuts, the signal cross-section is mostly independent of $m_\phi$ for $0.5 ~\text{GeV} < m_\phi < 20~\text{GeV}$ and is estimated to be $\sigma_{\text{sig}} \approx 2\kappa_{S/A}^2~\text{pb}$ while the background is $\sigma_{\text{bkg}} \approx 20~\text{pb}$ at $\sqrt{s} = 13~\text{TeV}$.\footnote{It is expected that these should also be approximately the expectation for the LHC Run 3 energy of $\sqrt{s} = 13.6~\text{TeV}$. } The high rate follows from the proton's large up-quark content compared to other flavors; coupling predominantly to other flavors would reduce the LHC cross-section, though those scenarios become more accessible to meson decay searches due to smaller CKM suppression.

\begin{figure*}[tbh!]
\centering
\includegraphics[width=0.49\linewidth]{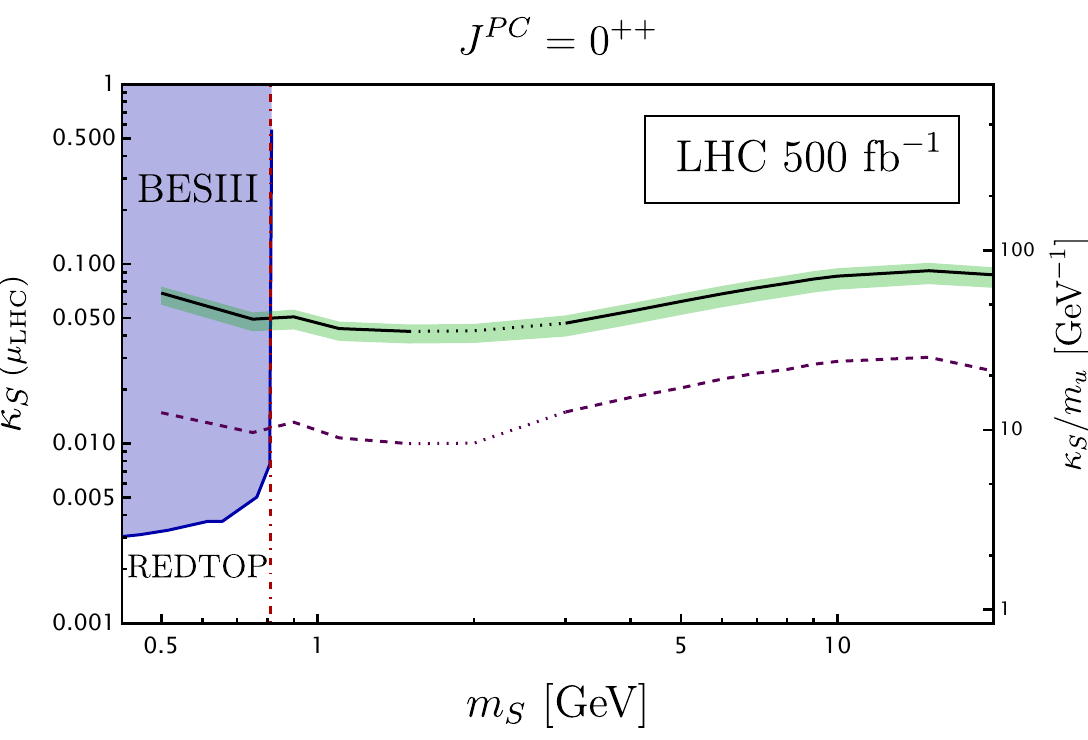}
\includegraphics[width=0.49\linewidth]{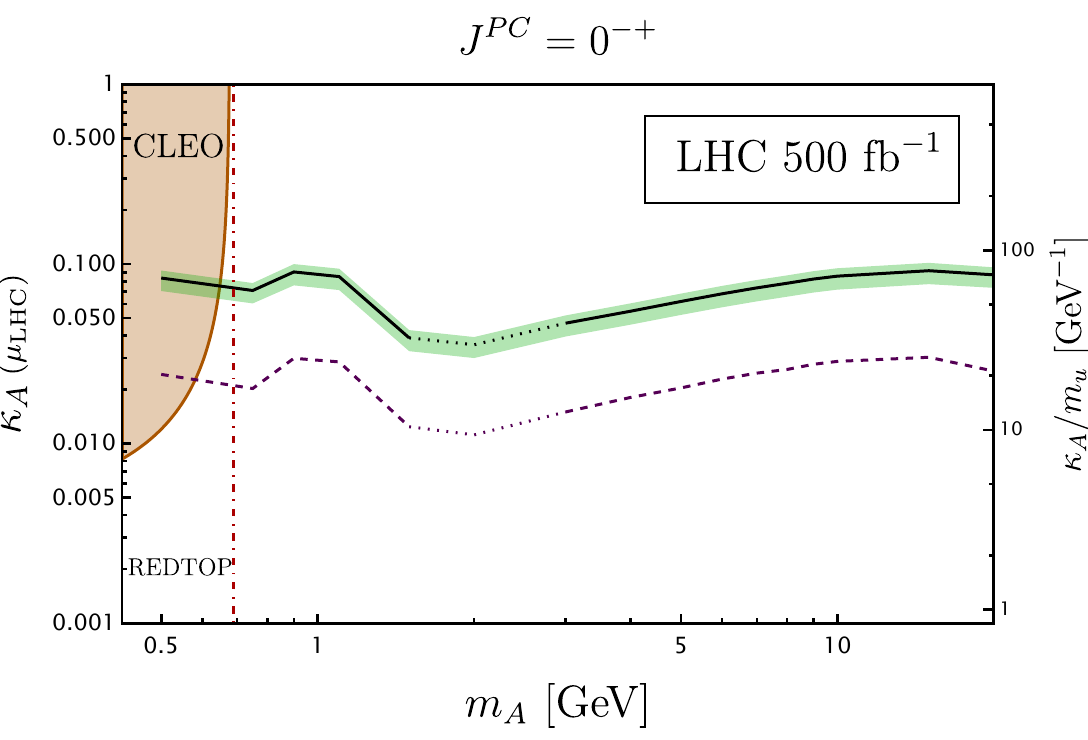}
\caption{Expected $95\%$ CL exclusion reach on the coupling $\kappa_{S/A}$ at the hard scale $\mu_{\rm LHC}=150$~GeV (left axis), equivalently on the renormalization-group-invariant ratio $\kappa_{S/A}/m_u$ (right axis), as a function of the scalar mass $m_{S/A}$ for P-even~(\textbf{left}) and P-odd~(\textbf{right}) hypothesis, estimated with 500 fb$^{-1}$ luminosity at $\sqrt{s} = 13~\text{TeV}$. The {\color{DarkGreen}{green}} band indicates the 50$\%$ background uncertainty and satisfies current trigger thresholds of $p_T>150~\text{GeV}$. The dashed {\color{Purple!75!Black}{purple}} curve shows projected reach with the same luminosity for a lower trigger threshold of $p_T>50~\text{GeV}$, which can be reached using dedicated trigger strategies. Existing constraints from BESIII ({\color{DarkBlue}{blue}}) and CLEO ({\color{Orange!75!Black}{orange}}) from exotic $\eta'$ decays are shown for comparison. The expected reach of the future REDTOP experiment for the same channels is shown as a {\color{DarkRed}{red}} dotted line. The LHC reach between 1.5 and 3 GeV is drawn as a dotted line in both panels, since in this region neither $\chi$PT nor the inclusive parton-shower description of the decay is under control and the reach is an interpolation. The BESIII, CLEO and REDTOP bounds are computed at the hadronic scale and have been run up under QCD to $\mu_{\rm LHC}$, which lowers them by $m_u(\mu_{\rm LHC})/m_u(2~{\rm GeV})=0.56$; the dashed curve measures $\kappa$ at $\mu\simeq50$~GeV and has been run to $\mu_{\rm LHC}$ as well, a $7\%$ effect.}
\label{fig:bound}
\end{figure*}

A key observable in this search is the jet mass. The standard jet mass is computed using all particle (or calorimetric) constituents clustered in the anti-$k_T$ jet and therefore captures both charged and neutral energy deposits. However, this quantity is sensitive to neutral pile-up contamination and underlying-event activity, which can significantly broaden the jet mass distribution of the signal. To address pile-up, we consider the jet mass of only the charged constituents (the ``track jet mass''). At the LHC experiments, charged tracks are able to be extrapolated to their production vertex, and charged tracks from simultaneous background collisions (pile-up) can be directly removed. The track-based observable is therefore significantly more robust against pile-up, particularly in high-luminosity conditions. While this strategy is expected to be necessary to make the analysis feasible, it loses the neutral component of the decays. These can be a large fraction of the decay for low masses, especially for the odd-scalar.

As the new particle is neutral, the decay always produces equal numbers of positively and negatively charged particles. This feature gives another handle in the signal and background discrimination, as the dominant background comes from quark jets, which carry net charge. We can use the Jet Charge $Q_k =(p_T^J)^{-k} \sum_i Q^i(p_T^i)^{k}$~\cite{Fermilab-Serpukhov-Moscow-Michigan:1979zgc,Berge:1980dx,EuropeanMuon:1984xji,ATLAS:2015rlw,Larkoski:2024uoc}, where $i$ runs for all the tracks in the reconstructed jet and $k$ is an optimized weighting factor. The optimal value for this search of $k$ is found to be 0, and we impose a cut enforcing $Q_0 = 0$, which follows the expectation for the signal to be balanced and clustered around a total charge of zero. This modifies the distributions from Fig.~\ref{fig:mass} by removing the events with an odd number of tracks since each track carries unit charge.

These observables are not perturbatively infrared and collinear safe. The track multiplicity changes under both soft and collinear emissions~\cite{Frye:2017yrw}; track jet mass can be in principle calculated but relies on non-perturbative track functions~\cite{Chang:2013rca,Chang:2013iba}, and $Q_0$ changes due to soft charged particles in a similar way as jet flavor is also IR unsafe~\cite{Banfi:2006hf,Czakon:2022wam}. Following the existing LHC measurements~\cite{ATLAS:2016vxz,ATLAS:2023dyu, ATLAS:2015rlw}, the IR is tamed by the detector resolution and the tracks are defined with $p_T>0.5$~GeV. For both the signal and background, their estimation is data-driven~\footnote{The estimation is data-driven by either fully constructed data-driven templates or the data used for the showering and hadronization fits.}, except for the region in parameter space where the signal is described by $\chi$PT.

We then perform a two-dimensional binning in the number of reconstructed tracks within the jet and the track jet mass. The signal populates the low-multiplicity and low-mass bins preferentially, whereas the QCD background exhibits a broader track multiplicity spectrum and a track jet mass distribution peaking at higher values. For each bin, we compute the expected number of signal and background events after the selection cuts described above and construct a Poisson log-likelihood. The test statistic is defined as the profile likelihood ratio, and we float the background by $50\%$ to estimate non-computed uncertainties. The reach of this search for 500~fb$^{-1}$ of data for both even and odd scalars can be seen in Fig.~\ref{fig:bound}. Most of the statistical reach comes from low track multiplicity, low mass bins, where the signal dominates over the background. A similar, but slightly weaker, reach can be obtained using simple cuts on these variables, highlighting that the important differentiator from QCD is the number of low-multiplicity, low-mass events. The LHC rate measures $\kappa$ at the hard scale of the process, $\mu_{\rm LHC}\simeq150$~GeV for the nominal selection, and Fig.~\ref{fig:bound} shows the reach at this scale, with the invariant ratio $\kappa/m_u$ on the right axis and the meson-decay bounds run up to $\mu_{\rm LHC}$ with the QCD running of Sec.~\ref{sec:model}. The nominal reach lies at $\kappa(\mu_{\rm LHC})\simeq0.04$--$0.1$, with $\kappa\simeq0.05$ at $m_\phi=1$~GeV, which corresponds to $\kappa/m_u\simeq35$--$80~{\rm GeV}^{-1}$, $m_u/\kappa\simeq13$--$30$~MeV and $\kappa(2~{\rm GeV})\simeq0.08$--$0.17$. The lowered trigger threshold discussed below measures $\kappa$ at $\mu\simeq50$~GeV, and the running between the two hard scales is a $7\%$ effect.

The reach can be further extended by relaxing the transverse-momentum requirements on the jet and the photon. The current choice of $p_T>150~\text{GeV}$ is imposed to satisfy existing trigger constraints rather than being optimal for the signal. Given the steeply falling parton luminosities, reducing the threshold to $p_T>50~\text{GeV}$ would lead to an enhancement of the production rate by a factor of $\sim60$. Such a reduction in thresholds would require dedicated trigger strategies, which are not currently implemented for $j + \gamma$. In particular, one could envision a ``Data Scouting''~\cite{CMS:2024zhe} or ``Trigger Level Analysis''~\cite{ATLAS:2025okg} implementation retaining information on the jet charge, track multiplicity, and the track jet mass in a $j+ \gamma$ topology computed during the online trigger analysis of the data. At $2\times10^{34}\,{\rm cm^{-2}s^{-1}}$, events with a photon candidate above 50~GeV at trigger level amount to a few kHz, dominated by misidentified jets, which is within the bandwidth of reduced-event-content streams \cite{CMS:2024zhe,ATLAS:2024xna}; events passing the full selection correspond to about $25\,{\rm Hz}$, and could be reduced further by requiring a recoiling jet with low track multiplicity at trigger level.

Since the signal systematically populates regions with low track multiplicity and small track jet mass, and these observables are computable in the trigger processing, a minimal amount of information needs to be stored to perform the analysis with the much lower thresholds available at the trigger level. We show the improved limits that could be obtained by lowering the trigger threshold to $50~\rm GeV$ in Fig.~\ref{fig:bound}, observing a gain of a factor of $\sim 4$ in the coupling strength reach (corresponding to an order of magnitude improvement in total signal strength), down to $\kappa(\mu_{\rm LHC})\simeq0.01$--$0.03$, i.e.\ $\kappa/m_u\simeq8$--$25~{\rm GeV}^{-1}$ or $\kappa(2~{\rm GeV})\simeq0.02$--$0.05$.

Another handle that can be used to further strengthen the signature is to perform additional analysis selections with higher $p_T$ cuts. While both signal and background cross sections fall with the higher $p_T$ cuts, the background jet distributions shift further towards higher multiplicity, with the signal independent of $p_T$. This can be used in case of a positive signature to further differentiate this hypothesis from standard QCD processes.

\subsection{Additional Constraints}

It is important to understand additional bounds that exist now or in the near future for this class of models. In the low mass region, below the reach of LHC experiments, beam dump searches, rare pion and kaon decay searches, astrophysical, and cosmological searches are very constraining~\cite{Batell:2018fqo,Batell:2021xsi}. We are primarily interested in masses above $500~\rm MeV$ where decays of heavier mesons, in particular $\eta^\prime$, can be important. Given the couplings in Eq.~(\ref{eq:eff_coupling}), new $\eta^\prime$ decay modes can occur. Using $\chi$PT, we compute the branching ratio for scalar plus neutral pion production to be~\cite{Batell:2018fqo}
\begin{equation}
{\rm Br}(\eta' \rightarrow \pi^0 S) \simeq 2.7\% \left( \frac{\kappa_S}{0.01}\right)^2 \times {\rm PS}_2 \, ,
\end{equation}
where ${\rm PS}_2$ is a two-body phase space function that tends to unity as $m_{\pi,S}/m_{\eta^\prime}\to 0$. This decay is kinematically allowed for $m_S<823~{\rm MeV}$. In this range, and for $m_S>279~{\rm MeV}$, $S$ decays dominantly into a pion pair. We conservatively place a bound in the $\kappa_S$ vs. $m_S$ plane shown in Fig.~\ref{fig:bound} by requiring that the branching ratio for $\eta^\prime\to \pi^0S(\pi^+\pi^-)$ not exceed the branching ratio for $\eta^\prime\to \pi^0\pi^+\pi^-$ of $0.36\%$ measured by BESIII~\cite{BESIII:2016tdb}. For a pseudoscalar with $m_A<679~{\rm MeV}$, we find the branching ratio
\begin{equation}
{\rm Br}(\eta' \rightarrow \pi^+\pi^- A) \simeq 1.6\% \left( \frac{\kappa_A}{0.01}\right)^2 \times {\rm PS}_3 \, .
\end{equation}
where ${\rm PS}_3$ is a three-body phase space factor that goes to one as $m_{\pi,S}/m_{\eta^\prime}\to 0$. For $m_A>414~{\rm MeV}$ it decays to three pions with a branching to $\pi^0\pi^+\pi^-$ of $40\%$~\cite{Bauer:2017ris,Aloni:2018vki} in this mass range. This decay mode, $\pi^02(\pi^+\pi^-)$, has been searched for at CLEO~\cite{CLEO:2008fxt} where the upper limit $1.8\times10^{-3}$ was set. We show the resulting limit in Fig.~\ref{fig:bound} by requiring that the rate for $\eta' \to \pi^+\pi^- A(\pi^0\pi^+\pi^-)$ not exceed this value. In Eqs.~(2) and~(3), as in Appendix~\ref{app:ChpTPythia}, the couplings enter through $\kappa/m_u$, and the numerical prefactors correspond to $m_u=2.13$~MeV, i.e.\ to $\kappa$ at $2$~GeV, with $\kappa(2~{\rm GeV})=0.01$ standing for $\kappa/m_u=4.7~{\rm GeV}^{-1}$ and $\kappa(\mu_{\rm LHC})=0.0056$ in Fig.~\ref{fig:bound}; the BESIII bound excludes $\kappa_S(2~{\rm GeV})\gtrsim0.005$--$0.02$ ($\kappa_S/m_u\gtrsim2.4$--$10~{\rm GeV}^{-1}$) for $m_S<0.8$~GeV and the CLEO bound $\kappa_A(2~{\rm GeV})\gtrsim0.015$--$0.1$ ($\kappa_A/m_u\gtrsim7$--$50~{\rm GeV}^{-1}$) for $m_A<0.7$~GeV.

These searches are based on samples of $\sim 10^4$-$10^6$ $\eta^\prime$ mesons. The proposed REDTOP experiment~\cite{REDTOP:2022slw} aims to collect up to $10^{12}$ $\eta^\prime$ mesons, which could increase the reach in the kinematically available regions substantially as shown in Fig.~\ref{fig:bound}. Additional improvement on the bounds could come from differential measurements.

Decays of heavy flavor mesons can also probe GeV-scale states coupled to quarks. However, an up quark specific coupling is particularly difficult to probe in $B^\pm$ decays. Tree-level two-body decays of $B^\pm$ mesons suffer from strong quark mixing suppression, with rates of $S$ or $A$ emission proportional to $\left|V_{ub}\right|^2\kappa_{S,A}^2\sim10^{-5}\kappa_{S,A}^2$. Decays that involve $\bar b\to\bar s c\bar c$ followed by a $c\bar c \to u\bar u$ rescattering and a further $\phi$ emission do not have the same CKM suppression but lead to many-body final states with rates that are difficult to reliably estimate. Simple arguments involving mixing with the pseudoscalar mesons along with measured SM $B$ decay rates indicate that current data could offer sensitivity to $\kappa_A\gtrsim0.5$ at $2$~GeV, i.e.\ $\kappa_A/m_u\gtrsim2\times10^2~{\rm GeV}^{-1}$ or $\kappa_A(\mu_{\rm LHC})\gtrsim0.3$, in the pseudoscalar case with the scalar case being less constrained. A more detailed study of these modes is beyond the scope of this work. Radiative decays of quarkonia, $J/\Psi$, $\Upsilon\to\gamma A$, $S$ are classic probes of light bosons~\cite{Wilczek:1977zn,McKeen:2008gd} but in this model occur through the effective one-loop coupling of the new bosons to two photons with branching fractions $\sim 10^{-8}\kappa_{A,S}^2$, below current limits. A comprehensive study of meson decays for this model could provide some sensitivity to masses below $\sim5~{\rm GeV}$; see, e.g.,~Refs.~\cite{Alda:2025uwo,Alda:2025nsz} for recent efforts along these lines.

Hadronically decaying states with masses of order tens of GeV or below can be difficult to constrain with data from colliders before the LHC; see, e.g.,~\cite{Evans:2018scg,Llorente:2018wup,Curtin:2018xsc}. Generally, they can appear as a rescaling of jet production cross sections, but difficult to directly probe without looking at detailed properties of the jets themselves, as we have proposed.

Lastly, we mention that the existence of new hadrophilic particles with masses of a few GeV can also affect the measurement of $\sigma(e^+ e^- \rightarrow \text{hadrons})$~\cite{Eidelman:1995ny}. The calculation is more under theoretical control at the Z pole. At one loop, it modifies the decay to u-quarks in a logarithmically UV-sensitive way, $\simeq-(\kappa^2/8\pi^2)\ln(M^2/M_Z^2)$, with $M$ the scale of the UV completion, which for $M\sim1$~TeV excludes $\kappa(M_Z)\gtrsim0.4$~\cite{ParticleDataGroup:2024cfk}, i.e.\ $\kappa/m_u\gtrsim3\times10^{2}~{\rm GeV}^{-1}$ with $m_u(M_Z)\simeq1.25$~MeV, which is $\kappa(\mu_{\rm LHC})\gtrsim0.4$ up to the small running between $M_Z$ and $\mu_{\rm LHC}$, well above the reach of Fig.~\ref{fig:bound}. More model independently, the real $\phi$ emission at the $Z$ pole, $\Gamma(Z\to u\bar u\phi)/\Gamma(Z\to u\bar u)\simeq3\times10^{-2}\kappa^2$ for $m_\phi=1$~GeV, acts in three-jet events as a scalar-gluon admixture at the level of $10^{-2}\kappa^2$ of the gluon rate, and the SLD limit on such an admixture~\cite{SLD:1996fuq} gives $\kappa(M_Z)\lesssim5$, i.e.\ $\kappa/m_u\lesssim4\times10^{3}~{\rm GeV}^{-1}$.  Similarly, it can also distort low-energy nuclei/pion scatterings~\cite{Meissner:2012ku}. A proper analysis for lepton colliders needs to be performed to have a clearer picture, where in most of these cases it is challenging to have theoretical control and disentangle QCD from potential BSM contributions.

\section{Conclusion}
\label{sec:conc}

\begin{figure*}[tbh!]
\centering
\includegraphics[width=0.49\linewidth]{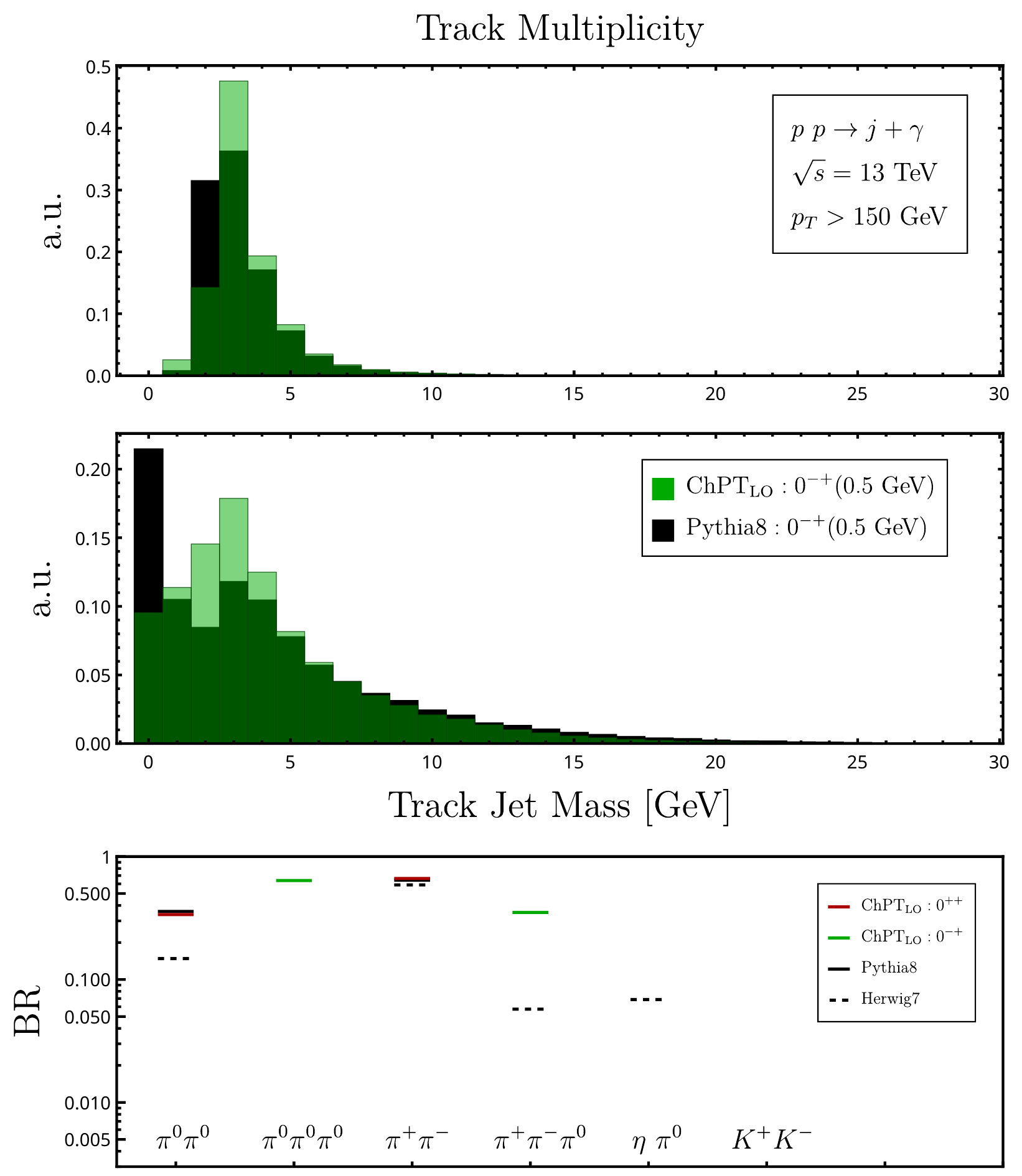}
\includegraphics[width=0.49\linewidth]{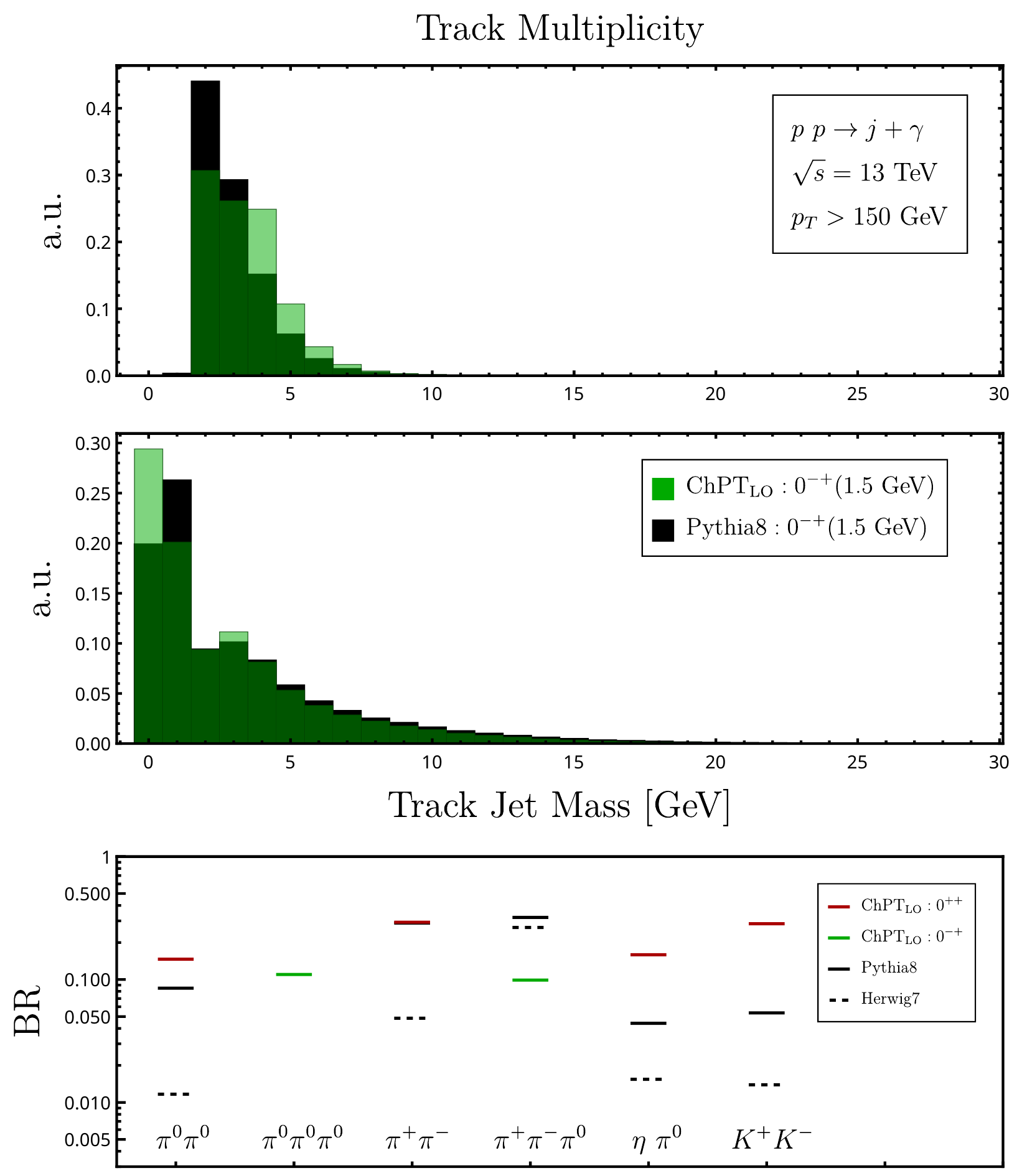}
\caption{Comparison between the leading order prediction of Chiral Perturbation Theory ($\chi$PT) vs \texttt{Pythia8.2} showering and hadronization. Normalized distributions for detector-level Track Multiplicity (\textbf{top}) and Track Jet Mass (\textbf{middle}). \textbf{Bottom:} Branching ratio for specific inclusive channels is shown. The solid black line shows the \texttt{Pythia8.2} prediction, the dashed black line shows the \texttt{Herwig7.2} prediction, while the $\chi$PT predictions for P-even and P-odd are shown in {\color{DarkRed}{red}} and {\color{DarkGreen}{green}}, respectively. \textbf{Left:} Assuming a scalar particle with 500 MeV mass. \textbf{Right:} Assuming a scalar particle with 1.5 GeV mass.}
\label{fig:sigvssig}
\end{figure*}

In this letter, we propose a search for GeV-scale particles that couple predominantly to light quarks using low-multiplicity jets at the LHC. Although such states can be produced at high rates, their prompt hadronic decays are typically assumed to be indistinguishable from those of QCD jets. We showed that this expectation is overly pessimistic. When produced in a hard process, the decay of a light gauge-singlet resonance is kinematically constrained. It leads to jets with unusually low charged-track multiplicity and small jet mass compared to QCD jets at the same transverse momentum.

We focus on the process $p p \rightarrow j + \gamma$ and construct a search based on the track multiplicity and the track jet mass. Using a simple detector simulation and conservative background uncertainties, we find strong sensitivity to scalar and pseudoscalar states in the mass range $0.5~\text{GeV}< m_{\phi} < 20~\text{GeV}$. Our results show that jets originating from light resonances produced in short-distance processes can exhibit distinctive properties even deep in the hadronic regime. In Appendix~\ref{app:datadriven}, we show that this analysis can use data-driven techniques, similar to those employed to tag quark- and gluon-initiated jets, to sidestep modeling uncertainties. Our results could be extended to other states, such as hadrophilic vector bosons. However, those models often suffer from anomalies that heavily constrain these constructions~\cite{Dobrescu:2014fca,Dror:2017ehi,Dror:2017nsg}.

\acknowledgments
We thank Brian Batell, Ayres Freitas, Andrew Larkoski, and Maxim Pospelov for helpful feedback on the draft and extend a special thanks to Andrew Larkoski for highlighting the importance of jet charge for this search. We also thank Benoît Assi, Yang Bai, Caterina Doglioni, David Morrissey, Katherine Pachal, and Isabel Trigger for fruitful discussions. This work is supported by Discovery Grants from the Natural Sciences and Engineering Research Council of Canada (NSERC). TRIUMF receives federal funding via a contribution agreement with the National Research Council (NRC) of Canada. 

\appendix

\section{Signal Uncertainties}\label{app:ChpTPythia}

The largest theoretical uncertainty in the signal prediction does not come from the production, but from the hadronic decay of the GeV-scale particle. In the mass window of $0.5~\GeV \lesssim m_\phi \lesssim 20~\GeV$, the decay interpolates between a regime where only a few channels are open and a regime where multi-hadron final states can be well described inclusively. For $m_\phi \lesssim 1.5~$GeV, only a limited number of final states are kinematically accessible. In this regime, we can use $\chi$PT to compute the decay amplitudes. For the pseudoscalar, we follow the approach of Refs.~\cite{Georgi:1986df,Bauer:2017ris,Bauer:2020jbp,Aloni:2018vki}, performing chiral rotations of $u$, $d$, and $s$ quarks to work in a basis where $A$ couples derivatively and does not mass mix with the $\pi^0$ (but does with $\eta^{(\prime)}$). We use this to compute the $A$ width, which for $m_A>3m_\pi$ is roughly
\begin{align}
\Gamma_{a\to3\pi}&=\Gamma_{a\to\pi^0\pi^+\pi^-}+\Gamma_{a\to3\pi^0} \nonumber
\\
&\simeq \frac{\kappa_A^2m_\pi^4 m_A}{768\pi^3 f_\pi^2(m_u+m_d)^2}\simeq0.02 \kappa_A^2m_A\, ,
\end{align}
which agrees with Refs.~\cite{Bauer:2017ris,Aloni:2018vki} but is about a factor of 10 smaller than what is found in Refs.~\cite{Murayama:2026ioh,DiLuzio:2026poh}. In this expression, we have used $m_u+m_d\simeq 7~{\rm MeV}$~\cite{FermilabLattice:2018est}. In the case of the scalar, we follow Ref.~\cite{Batell:2018fqo}, with $S$ appearing as a chiral-symmetry--breaking spurion in the quark mass matrix.

An important distinction between $A$ and $S$ is the number of pseudoscalar mesons that they produce when they decay. At leading order in chiral-symmetry breaking, $S$ decays to pairs of mesons while $A$ decays to three mesons. The analysis we performed is strongly sensitive to the number of charged tracks from the decay, which is the main difference between the reaches for $A$ and $S$ seen in Fig.~\ref{fig:bound}. 

Above $\sim 1.5~$GeV, many hadronic channels open, and an inclusive description of the decays becomes increasingly accurate. In this region, the detailed exclusive decay structure is less important, and a parton-shower description of an on-shell $\bar u u$ pair provides a reasonable approximation. We can see the comparison between the jet properties and exclusive decay channels from different approaches in Fig.~\ref{fig:sigvssig}. We note that while the jet properties are similar up to $\sim 1.5~$GeV, the exclusive channel branchings are not well described by current showering and hadronization tools. 

Because the jet properties are more universal, and mostly only sensitive to the mass of the scalar, we approach the signal simulation as follows: We use explicit exclusive decays as computed in $\chi$PT for masses below $1.5~$GeV and switch to \texttt{Pythia8.2}/\texttt{Herwig7.2} showering and hadronization above that scale. The precise value of the transition scale is not physical and introduces a small modeling uncertainty. 

Above a few GeV the signal is the hadronization of a color-singlet $u\bar u$ system of mass $m_\phi$, which is the same system produced in $e^+e^-\to u\bar u$ at $\sqrt{s}=m_\phi$, so that multiplicity and event-shape measurements at low-energy $e^+e^-$ colliders can in principle validate the generators in this regime. The tunes of \texttt{Pythia8.2} and \texttt{Herwig7.2} are, however, dominated by LEP data, and the region between 1.5 and 3 GeV, where neither $\chi$PT nor the parton shower is under control, is treated as an interpolation and shown as a dotted line in Fig.~\ref{fig:bound}.

For this work, we only consider $u$ quark-initiated $\phi+\gamma$ production since the gluon-initiated production through the effective $\phi GG$ operator is loop-suppressed and suppressed at high transverse momenta. This $\phi GG$ operator is relevant when considering decays of heavy flavor mesons as well as the decays of the new scalars above $\sim 1.5~{\rm GeV}$~\cite{Alda:2025uwo,Alda:2025nsz,Gieseke:2025gfq}.

\section{Background Uncertainties and Data-Driven Templates Using DiJets}\label{app:datadriven}

In this appendix, we discuss the background estimation and uncertainty determination of the QCD $j + \gamma$ process. The parton-level cross-section can be systematically improved; however, a large source of uncertainty remains from the non-perturbative showering and hadronization. This makes \textit{ab initio} simulation of $j + \gamma$ challenging. 

The differences between showering and hadronization can be explored by comparing two different approaches. We compare, in Fig.~\ref{fig:HWPY}, the reconstructed jet properties at leading order of production from showering and hadronization performed by \texttt{Pythia8.2} and \texttt{Herwig7.2}. While we use \texttt{Pythia8.2} in the main paper, the bounds derived from \texttt{Herwig7.2} are statistically equivalent once the background uncertainty is included. 

The difficulty lies in obtaining a reliable estimate of the expected background events that are similar to the signal. This kinematic region is far away from where the fits for showering and hadronization tools are performed. For example, the production of QCD resonances with the same quantum numbers as the scalar (such as $f_0(1370),~f_0(1500),~f_0(1710) $, and so on) or pseudoscalar ($\eta(1285),~\eta(1405),~\eta(1475)$, etc.), with few additional neutral mesons can look like the signal.

In the showering and hadronization performed with \texttt{Herwig7.2}, these anomalous low-track jets are present, which explains the deformation for lower track multiplicity and track jet mass seen in Fig.~\ref{fig:HWPY}. Nevertheless, it is difficult to know how reliable the estimation of those jets is. Their precise determination is crucial for a reliable search for BSM particles with similar characteristics.

Thankfully, there are more robust ways to significantly improve background estimation without relying on theoretical uncertainties. One way to estimate the background shape is to infer the underlying distributions from similar processes. Specifically, extracting jet templates from signal-poor samples and then using them in the analysis. This method is routinely used in LHC searches to produce quark and gluon jet templates~\cite{Gallicchio:2011xq,Larkoski:2019nwj,ATLAS:2014vax,ATLAS:2023dyu}. This approach bypasses non-perturbative theoretical uncertainties and constructs data-to-data jet templates that would account for the correct fraction of those rare low multiplicity background jets. 

This background extraction can be performed using the two-jet final state, even in the presence of this class of BSM models. Consider the process $pp \rightarrow j j$ where the jet is either from QCD or from the scalar $\phi$. The signal process is similar to the $j+\gamma$ case, with a slightly higher parton-level cross-section. On the other hand, the QCD background this time is significantly larger at the parton level, mostly induced by the number of possible parton combinations and the removal of the $\alpha_\mathrm{EM}$ vertex. This final state provides a background-rich sample that can be used to construct pure-quark and pure-gluon jet templates, following the methods of the standard template extractions.

To show the feasibility, we extract simulation-to-simulation templates using the matrix method~\cite{ATLAS:2017nma}. The procedure closely follows Ref.~\cite {ATLAS:2017nma}, with the additional verification of signal contamination. We produce $pp \rightarrow j j$ samples with \texttt{MadGraph5\_aMC@NLO} + \texttt{Pythia8.2}, then consider only balanced back-to-back jets (both with $p_T>150$ GeV) which are either central ($\eta<1.4$) or forward ($1.5<\eta< 2.4$). The quark and gluon content of these two sets is different and is mostly a perturbative quantity, which can be estimated using \texttt{Pythia8.2}. Additionally, the signal cross-section after showering and detector effects is negligible, with values of around 7 fb in the central region and 2 fb in the forward region for $m_{\phi} = [0.5,20]$ GeV and $\kappa_{S/A}=1$. At the same time, the background is significantly higher, with an expected cross-section of 8228 pb in the central region and 1651 pb in the forward region. This means that in the few fb of data available~\footnote{Depending on the jet $p_T$, data-driven template methods use various pre-scaled jet triggers that randomly select a fraction of the data passing selections for recording, leading to substantially reduced datasets compared to the full LHC run. Given the very high QCD cross-section, these reduced datasets are in many cases still sufficient for studies of QCD properties.} for the background extraction, there is virtually no signal contamination. Throughout this process, careful attention must be paid to the kinematic properties of the background estimate to ensure it properly mimics the true $\gamma$+jet distribution, and that sufficient statistics are available at low charged-track multiplicity in the dijet sample to yield a statistically meaningful background estimate.

As this is a simulation-to-simulation extraction, the process is extremely efficient, even in the presence of underlying events. This can be seen in Fig.~\ref{fig:HWPY}, where we compare the template extracted from the \texttt{Pythia8.2} di-jet sample to the simulated jet from $j+\gamma$.  The templates are created for different $p_T$ bins with the perturbative prediction for the quark and gluon templates at their respective ranges. Once the template is obtained, we have applied it to the $j+\gamma$ process and estimated the quark and gluon content. Because the procedure closes out nearly perfectly, we use the distributions obtained directly from \texttt{Pythia8.2} for the analysis presented in the paper.

\bibliography{bibmulti}

\end{document}